\documentclass{emulateapj}
\pdfoutput=1
\usepackage{amssymb}
\usepackage{gensymb}
\usepackage{graphicx}
\usepackage{amsmath}
\usepackage{wasysym}
\usepackage{verbatim}

\begin{document}
\title{Explaining TeV Cosmic-Ray Anisotropies with Non-Diffusive Cosmic-Ray Propagation}
\author{J.~Patrick~Harding}
\affil{P-23, Los Alamos National Laboratory, Los Alamos, NM 87545, USA} 
\email{jpharding@lanl.gov} 

\author{Chris~L.~Fryer}
\affil{CCS-2, Los Alamos National Laboratory, Los Alamos, NM 87545, USA} 
\email{fryer@lanl.gov} 

\and

\author{Susan~Mendel}
\affil{ISR-2, Los Alamos National Laboratory, Los Alamos, NM 87545, USA} 
\email{smendel@lanl.gov} 


\begin{abstract}
  Constraining the behavior of cosmic ray data observed at
  Earth requires a precise understanding of how the cosmic rays
  propagate in the interstellar medium. The interstellar medium is not
  homogeneous; although turbulent magnetic fields dominate over large
  scales, small coherent regions of magnetic field exist on scales
  relevant to particle propagation in the nearby Galaxy. Guided
  propagation through a coherent field is significantly different from
  random particle diffusion and could be the explanation of spatial
  anisotropies in the observed cosmic rays. We present a Monte Carlo
  code to propagate cosmic particle through realistic magnetic field
  structures. We discuss the details of the model as well as some
  preliminary studies which indicate that coherent magnetic structures
  are important effects in local cosmic-ray propagation, increasing
  the flux of cosmic rays by over two orders of magnitude at
  anisotropic locations on the sky. The features induced by coherent
  magnetic structure could be the cause of the observed TeV cosmic-ray
  anisotropy. 
  \end{abstract}



\maketitle
\section{Introduction}
At high energies, cosmic ray particles are one of the most important
handles for understanding the Galaxy; the energy density of cosmic
rays is similar to that of Galactic magnetic fields and
starlight. However, the sources of the cosmic rays remain a
mystery. Many sources, from active galactic nuclei and supernovae to
dark matter annihilation have all been postulated as the origin of the
measured cosmic rays. Because cosmic rays are charged, they do not
point back to their sources, so in order to understand their creation,
one must understand the process of cosmic-ray propagation. Local
measurements of cosmic rays have shown several signatures which cannot
be explained by standard astrophysical propagation models. In order to
understand the significance of these signatures and the sources of
these cosmic rays, we propose a new framework for modeling cosmic-ray
propagation.

Current propagation codes, such as GALPROP~(\cite{Strong:1998pw}),
tend to focus on the diffusive regime of particle propagation,
considering particle diffusion and losses in the regime where the
magnitude of the magnetic field is relevant but not the direction,
i.e. the diffusion limit.  For particles which propagate over large
regions of space ($\gtrsim100{\rm\,pc}$), the magnetic field may be
dominated by a stochastic, chaotic component which does have this
behavior~(\cite{Giacinti:2011mz}). However, even in these regions, the
strength of magnetic turbulence is typically smaller than that of the
guiding field, so the coherent field may still play an important
role~(\cite{2011Natur.478..214G}). This can be seen in studies of
pitch-angle scattering in large interstellar coherent magnetic
structures~(\cite{1984ApJ...284..817B,2014ApJ...791...51D}) and in
studies of parallel and perpendicular
diffusion~(\cite{1999ApJ...520..204G,Tautz:2013vba,Tautz:2014ega,2015JGRA..120.4095H,2015AdSpR..56.1264S}). We
plan on incorporating these cross-fieldline transport effects into our
framework in the future.

Over propagation distances smaller
than those considered in these papers, such as those relevant for cosmic rays observed at
energies above the TeV scale, the magnetic fields can also have strong
spatial correlations; the magnetic fields which affect the local
galaxy can be coherent, as opposed to the chaotic, stochastic magnetic
fields believed to dominate throughout the galaxy on larger
scales~(\cite{2014ASPC..484...42F,2015JPhCS.577a2010F,2015ApJ...814..112F}).
Some indications of these coherent magnetic structures, which can have
correlation length of tens of parsecs, have been observed, such as
those from the heliotail or the recently-observed features by
IBEX~(\cite{2014Sci...343..988S}) (though the features in that paper are given in terms of a diffusive model). It is the nondiffusive regime of
cosmic-ray propagation that we consider here.

When particles see these small-scale magnetic structures, their
propagation is much different than in the standard chaotic
fields. Using a combination of discrete diffusion Monte Carlo and
direct Monte Carlo, we simulate cosmic-ray propagation including both
the magnetic field strength and direction. By including the magnetic
field direction, we are able to account for coherent features in the
magnetic field and correlations between nearby regions in magnetic
field strength.
\section{Physical Motivation}

\begin{figure}[t]
\begin{center}$
\begin{array}{c}%
\includegraphics[width=0.95\columnwidth]{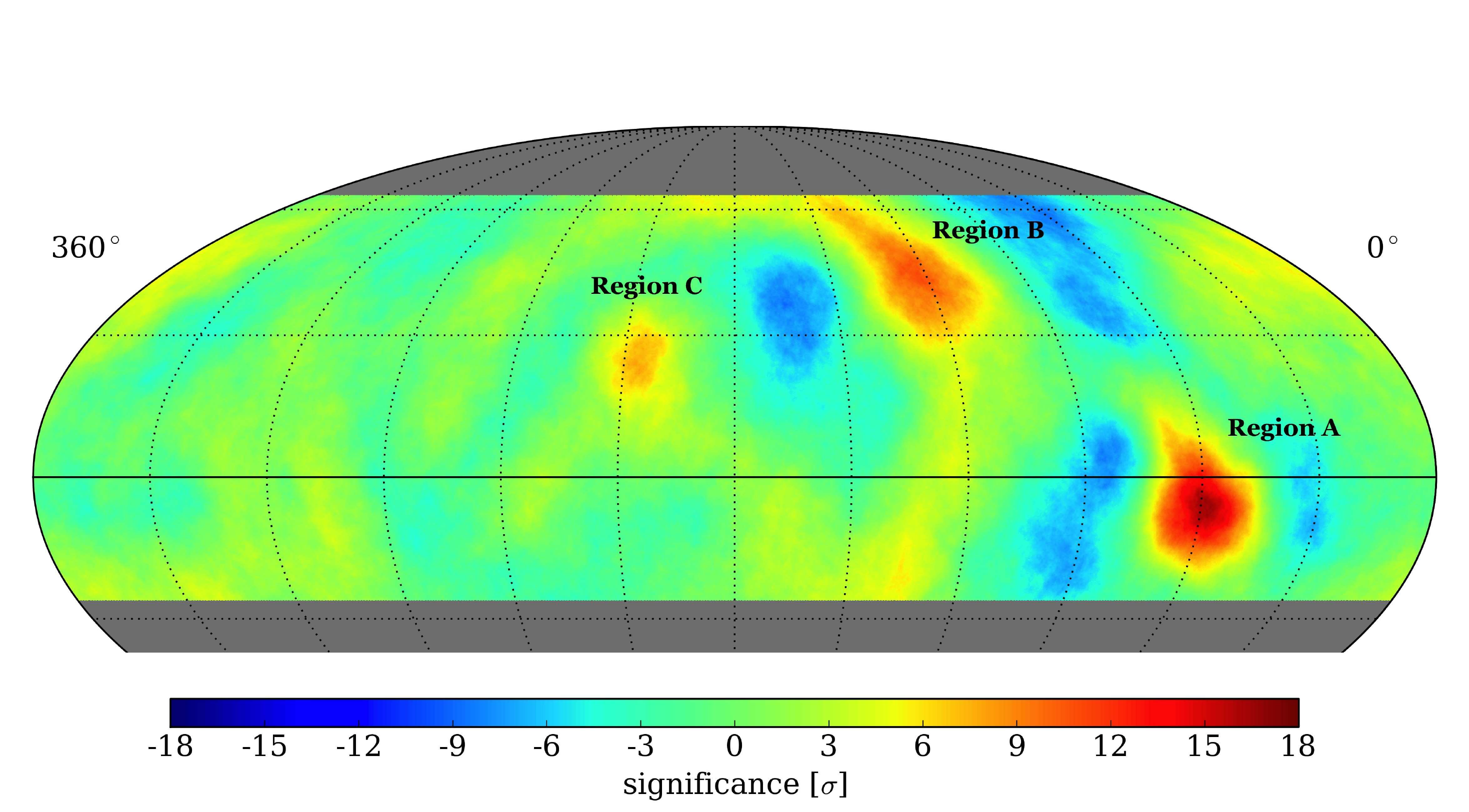}
\end{array}$
\end{center}
\caption{\small The observed anisotropy of hadronic cosmic rays as
  seen by HAWC~(\cite{2014ApJ...796..108A}). Three regions of significant excess
  over the background (Regions A, B, and C) are observable on scales
  of less than ten degrees on the sky.
\label{CRanisotropy}}
\end{figure}
One particular observation that has motivated this propagation code
is the TeV cosmic-ray anisotropy. Anisotropies in the arrival directions of TeV cosmic rays (CRs) have been observed
by many
experiments, both on large angular scales~(\cite{Amenomori:2005dy,Amenomori:2006bx,Guillian:2005wp,Abdo:2008aw,2009ApJ...692L.130A,Zhang2009,2010ApJ...711..119A,2010ApJ...712.1100M,2010ApJ...718L.194A,deJong:2012hk,2011ICRC....1....6C,2013ApJ...765...55A,2015ApJ...809...90B}) 
and small angular scales~(\cite{Amenomori2007,Abdo:2008kr,2011ApJ...740...16A,Aartsen:2012ma,DiSciascio:2013cia,ARGO-YBJ:2013gya,2014ApJ...796..108A}). The
observed small-scale cosmic-ray anisotropy from the High Altitude Water Cherenkov
(HAWC) collaboration~(\cite{2014ApJ...796..108A}) is shown in
figure~\ref{CRanisotropy}. While a large-scale dipole anisotropy is expected from asymmetries in source distributions, the source of the small-scale anisotropic CRs remains a
mystery. It appears that The small-scale anisotropy may be made up mostly of
hadrons~(\cite{Abdo:2008kr}), but the gyroradius of a TeV proton in the local magnetic
field is only a few milliparsecs, which is much closer than any
currently-known candidate source of CRs. Even the neutron decay length
of 0.1 parsecs is thousands of times smaller than the nearest known
supernova remnant, the Geminga pulsar. The energy spectrum of the
anisotropic CRs is also significantly harder than the spectrum of
isotropic CRs.

Production of a localized hadronic region of CRs requires anisotropic
CR propagation, as CRs from any source further than 0.1 parsec from
the Earth are made nearly isotropic through diffusive propagation to
the Earth. Several sources of the CR anisotropy have been considered,
including magnetic mirrors~(\cite{Drury:2008ns}), anisotropic
turbulence~(\cite{Malkov:2010yq}), a local source in the
heliotail~(\cite{Lazarian:2010sq}), strangelets~(\cite{Kotera:2013mpa}),
turbulent mixing from the large scale
anisotropy~(\cite{Giacinti:2011mz}), and nearby dark matter
subhalos~(\cite{Harding:2013qra}). Most of these explanations, however,
require anisotropic cosmic-ray propagation. Many of these
explanations of the cosmic-ray anisotropy require limited particle
diffusion, such as that through a local coherent magnetic field
structure. By including this magnetic field structure in the cosmic
ray propagation, we propose to test several of the source populations
and magnetic field configurations that can explain the TeV cosmic-ray
anisotropy and quantify which features can reproduce the observed
anisotropy.

Another relevant measurement which cannot be explained with standard
astrophysical cosmic rays is the observation of an excess in local
cosmic-ray
positrons~(\cite{Adriani:2011xv,FermiLAT:2011ab,Aguilar:2013qda}). The
origin of these positrons is currently unknown, with explanations
ranging from nearby pulsars~(\cite{Linden:2013mqa}) to exotic
explanations like dark matter
annihilation~(\cite{Cholis:2013psa}). However, the questions remain of
how far the source of these positrons is from the Earth and the
spectrum of their source, both of which are intimately tied to the
propagation of the positrons in the local neighborhood.
\begin{figure}[t]
\begin{center}$
\begin{array}{c}
\includegraphics[width=0.95\columnwidth]{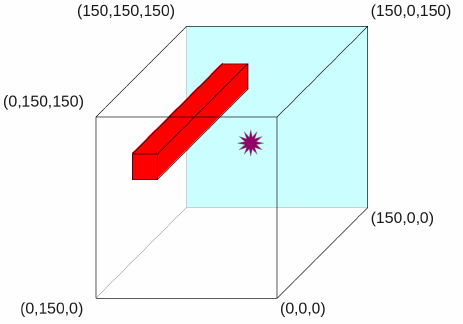}
\end{array}$
\end{center}
\caption{\small A schematic of the source, located at
  $(x,y,z)=(75{\rm\,pc},75{\rm\,pc},75{\rm\,pc})$ (magenta star) and
  the coherent magnetic field $a_{\rm min}<y,z<a_{\rm max}$ (red
  tube). The $x=150{\rm\,pc}$ wall (in blue) is our proxy for the
  Earth sky. Note that in our figures of the $x=150{\rm\,pc}$ wall,
  the coordinates of the y-axis go from left-to-right instead of
  right-to-left.
\label{BfieldGeometryFig}}
\end{figure}
\section{The Geometry of the Code}
This code is designed for propagation of highly relativistic
particles, which move near to the speed of light. Specifically, the
code is designed to work only for particles with $v\approx c$ and will
finish propagation in the case that the particles energy drops low
enough to make the particle speed appreciably lower than $c$. These
particles travel long distances before losing an appreciable amount of
energy. The code is currently set up to work with cosmic-ray
electrons, positrons, protons, and ions, but here we only consider
cosmic-ray protons, which are expected to be the dominant form of
cosmic rays with TeV-PeV energies creating the observed cosmic-ray
anisotropies.

The code is based on a 3-dimensional grid of cells, with the physical
properties constant within each cell. The cells are set to be much
larger than the scattering length of each cosmic ray and much larger
than the Larmor radius of the particle. Each cell has several
properties which are allowed to vary cell-to-cell, including the
particle density, particle temperature, hydrogen and helium fraction,
turbulent magnetic field strength, scale of the turbulent magnetic
field coherence length, and coherent magnetic field strength and
direction. Each simulated cosmic ray is then propagated through these
cells. Characteristic values for these parameters are shown in
table~\ref{paramstab}. For the calculations in this paper, we consider
cubic cells with 3\,pc spacing, though any spacing which is much larger than the particle's mean-free-path (for chaotic fields) and larger than the particle Larmor radius (for coherent fields) could be used. The walls of the simulation act as
absorption boundaries, with particle propagation ending once the
particle hits one of these bounding walls.

For our first simple geometry, we consider two regions, one outside of
the coherent field and one within the coherent field. Our coherent
field region passes along the x-direction between $a_{\rm
  min}<y,z<a_{\rm max}$. For a schematic, see
figure~\ref{BfieldGeometryFig}. For the simple case shown here, we
assume that there is no coherent field outside of this region, only
turbulent fields. We also assume there is no turbulent field within
this region, only coherent fields. This neglects the effects of
scattering within regions with coherent fields, such as particle
feedback within the coherent region producing a small turbulent
magnetic field. It is therefore not a definitive study of the
geometries which will produce the observed anisotropy, but rather a
test geometry for the code which indicates that such coherent fields
could be important for studies of cosmic ray anisotropies. How the
inclusion of turbulent components with coherent components will affect
the results will be studied in a follow-up paper. However, the authors
believe that the qualitative behavior will not change and that the
quantitative behavior should be similar. This is because many of the
particles which may scatter out of the coherent-dominated region will
scatter back into it. Also, the difference in timescales between
turbulent and coherent motion (linear versus square root in time)
means that particles will travel a long way along coherent streams
before scattering out of the coherent-dominated region.

The source distribution allows for multiple sources with varying
spectra and particle-type production. For simplicity, however, herein
we consider a single source located at
$(x,y,z)=(75{\rm\,pc},75{\rm\,pc},75{\rm\,pc})$ in the center of our
simulation region. Our simple particle spectrum is 100\% protons all
of which start with $10{\rm\,TeV}$ of energy.
\begin{table}
\begin{center}
\begin{tabular}[t]{|c|c|c|c|c|c|c|c|}
  \hline
  $n$ & $n_e$ & $k_BT$ & $x_{\rm H}$ & $x_{\rm He}$ & $\lambda_{\rm max}$ & $b_{\rm t}$ & $\lvert\vec{B}_{\rm coh}\rvert$\\
  \hline
  $0.01$ & $0.01$ & $10^{-4}$ & 0.9 & 0.1 & $10^{15}$ & $0{\rm\,\mu G},$ & $0{\rm\,\mu G},$\\
  ${\rm\,cm}^{-3}$ & ${\rm\,cm}^{-3}$ & ${\rm\,MeV}$ & & & ${\rm\,cm}$ & $3{\rm\,\mu G}$ & $3{\rm\,\mu G}$\\
  \hline
\end{tabular}
\caption[Table of the characteristic values for cell parameters.]{
Table of the characteristic values for the parameters in the cells. The parameters are, in order: the combined number density of neutral hydrogen and helium, the number density of ionized electrons, the particle temperature, the hydrogen fraction, the helium fraction, the turbulent magnetic field scale length, the maximum turbulent field strength, and the coherent field strength. These parameters are roughly equivalent to the values in the local bubble~(\cite{1995SSRv...72..499F}). For cells outside of our coherent field, the coherent field strength is $0.0{\rm\,\mu G}$ and the turbulent field strength is $3{\rm\,\mu G}$~(\cite{Kistler:2012ag}). Inside our simple coherent field, the turbulent field strength is $0.0{\rm\,\mu G}$ and the coherent field strength is $3{\rm\,\mu G}$ along the x-direction.
\label{paramstab}}
\end{center}
\end{table}
\section{Particle Propagation}
The propagation of particles through the intergalactic medium involves
a variety of particle processes, including particle-particle
collisions and interaction of charged particles with magnetic fields,
which affect the trajectory of the particle's path and energy
losses. Here we discuss an overview of how the code deals with
particle propagation, with further details in later sections.

At the beginning of the transport, the particle energy, starting
location, type, and direction are determined randomly from the input
sources. The particle is looped over coarse timesteps. For each coarse
timestep, the cell position of the particle is found, along with the
parameters for that cell. If there is a coherent magnetic field in the
cell, then the transport Monte Carlo propagates the particle until
either the coarse timestep is over or the particle leaves the cell. If
there is no coherent magnetic field in the cell, then the discrete
diffusion Monte Carlo is used to propagate the particle until either
the coarse timestep is over or the particle leaves the cell. After
each timestep, the energy losses for the particle over this time are
calculated. This process is repeated over timesteps and cells until
the particle exits the simulation.

It is important to note that while the studies in this paper consider
coherent and diffusive magnetic fields separately, in all regions in
space both of these effects contribute to the particle
motion. Additionally, there is not a distinction between
``single-particle'' motion and ``ensemble diffusive'' motion in the
trajectories of true particles. Even for regions of space in which the
ensemble behaves diffusively, individual particles actually have
deterministic trajectories. Particularly over distances of less than a
single scattering length, these deterministic propagation effects can
be seen. This is part of the purpose of this propagation code - to
show the effects which are often neglected in diffusion-only codes.
\subsection{Particle Transport Monte Carlo}\label{PTMC}
The particle transport Monte Carlo follows each particle as it
traverses each cell and passes into the adjacent cell. Within each
cell, chaotic fields are assumed to be coherent over their small
coherence length, so over these short distances the particle motion
can be solved exactly. Particles within a magnetic field of strength
$B$, energy $E$, and charge $Ze$ have a Larmor radius
\begin{equation}
r_L=1.1\times10^{-9}{\rm\,pc^{-1}}Z^{-1}(E/{\rm MeV})(B/{\rm \mu G})^{-1}\enspace.
\end{equation}
The equation of motion for this particle is~(\cite{Kistler:2012ag})
\begin{equation}
\frac{\mathrm{d}\vec{v}}{\mathrm{d}t}=\frac{c}{r_L}\vec{v}\times\hat{B}\enspace,
\end{equation}
where $\hat{B}$ is the direction of the net magnetic field (coherent
and turbulent) and $\lvert\vec{v}\rvert=c$. For our calculations, we 
choose time in years and distance in parsecs 
($c\approx0.31{\rm\,pc/yr}$).

The equation of motion can be solved analytically, over the time in
which the particle traverses a single coherent region of the turbulent
magnetic field. This is most easily done in a coordinate system with
an axis along the magnetic field. For our calculations, we use the
coordinates $(\hat{i},\hat{j},\hat{k})$ with
\begin{align}
\hat{j} &= \hat{B} = {\begin{pmatrix} \frac{B_x}{B} && \frac{B_y}{B} & \frac{B_z}{B}\end{pmatrix}}\label{hatj}\\
\hat{k} &= \frac{\hat{v_0}\times\hat{B}}{\lvert\hat{v_0}\times\hat{B}\rvert}\nonumber\\
 &= \begin{pmatrix} \frac{v_{0y}B_z-v_{0z}B_y}{\lvert\vec{v_0}\times\vec{B}\rvert} & \frac{v_{0z}B_x-v_{0x}B_z}{\lvert\vec{v_0}\times\vec{B}\rvert} & \frac{v_{0x}B_y-v_{0y}B_x}{\lvert\vec{v_0}\times\vec{B}\rvert} \end{pmatrix}\label{hatk}\\
\hat{i} &= \hat{j}\times\hat{k}=\hat{B}\times\frac{\hat{v_0}\times\hat{B}}{\lvert\hat{v_0}\times\hat{B}\rvert}\nonumber\\
 &= {\begin{pmatrix} \frac{B^2 v_{0x}-B_x(\vec{v_0}\cdot\vec{B})}{B\lvert\vec{v_0}\times\vec{B}\rvert} & \frac{B^2 v_{0y}-B_y(\vec{v_0}\cdot\vec{B})}{B\lvert\vec{v_0}\times\vec{B}\rvert} & \frac{B^2 v_{0z}-B_z(\vec{v_0}\cdot\vec{B})}{B\lvert\vec{v_0}\times\vec{B}\rvert}\end{pmatrix}}\label{hati}\enspace,
\end{align}
with $\hat{v_0}$ the initial direction of the particle velocity. We
refer to the j-component of the initial velocity (along the direction
of $\vec{B}$) as $v_{\|}$ and the magnitude of the initial velocity
perpendicular to the magnetic field as $v_{\bot}$
($v_{\bot}=\sqrt{\vec{v_0}^2-v_{\|}^2}$). Note that at $t=0$, the initial velocity is entirely along the $\hat{i}$ and $\hat{j}$ directions. After a time $t$, the velocity is
\begin{eqnarray}
v_i &=& v_{\bot}\cos(c t/r_L)\\
v_j &=& v_{\|}\\
v_k &=& v_{\bot}\sin(c t/r_L)\enspace.
\end{eqnarray}
To get the results in the original $(x,y,z)$-coordinates, we need to project these values back into the $(x,y,z)$-space, using the components given in equations~\ref{hatj}-\ref{hati}. 

A similar calculation to the particle velocity can be done to
calculate the particle position as well. However, for problems in
which the Larmor radius is much smaller than the cell size and the
mean-free-path, the motion orthogonal to the magnetic field can be
neglected and the equation simplifies. In this case, which applies to
the geometry we consider herein, the particle follows the magnetic
field with a velocity
\begin{equation}
\vec{x}(t)=\vec{x}_0+v_{\|}t\hat{B}=\vec{x}_0+\frac{\vec{v_0}\cdot\vec{B}}{B}t\hat{B}\enspace.
\end{equation}
In this case, however, the full path length traversed by the particle,
which is needed to account for absorption, is given by $ct$.  Note
that we consider the particle's energy to remain constant within each
cell, which is a reasonable approximation for the short distances and
high energies we consider here. We also have assumed ultrarelativistic
particles, with $v\approx c$ during the propagation. For low-energy
particles or extremely large cells, these approximations may no longer
be valid.

The particle transport in regions of a coherent magnetic field is
driven by taking steps over which the turbulent magnetic field is
constant~(\cite{Petrosian:2004ft}). Even though turbulent magnetic
fields have small coherence lengths, over these lengths they can be
thought of as coherent. We determine the distance that a particle
travels before it encounters a change in the turbulent magnetic field
following~\cite{Fryer:2006wy}:
\begin{align}\label{ksc}
&k_{\rm sc}=1/\lambda_{\rm sc}\nonumber\\
&=4.7\times10^{12}{\rm\,pc^{-1}}\left(\frac{\lambda_{\rm max}}{\rm cm}\right)^{-1/2}\left(\frac{B_{\rm t}}{\rm \mu G}\right)^{1/2}\left(\frac{E_{\rm prot}}{\rm MeV}\right)^{-1/2}
\end{align}
where $\lambda_{\rm max}$ is the scale length of the turbulent field,
$B_{\rm t}$ is the amplitude of the turbulent magnetic field, and
$E_{\rm prot}$ is the proton energy. Note that $B_{\rm t}$, the
instantaneous amplitude of the turbulent magnetic field, should not be
confused with the maximum amplitude of the turbulent magnetic field
$b_{\rm t}$.

We also consider the possibility of the particle being absorbed into
the ISM, which provides an additional opacity. This is due to
scattering on nuclei and is proportional to the total number density
of nuclei (not nucleons) to scatter with. We parameterize this
absorption using~(\cite{Fryer:2006wy})
\begin{equation}\label{kabs}
k_{\rm abs}=1/\lambda_{\rm abs}=9.9\times10^{-8}{\rm\,pc^{-1}}(n/{\rm cm^{-3}})\enspace.
\end{equation}
This accounts for all particle absorption on the ISM. This absorption
is determined solely by the number density of ISM particles within
each cell and therefore does not change until the particle leaves the
cell.  In the ISM $k_{\rm sc}>>k_{\rm abs}$, though in general this
could change for high-density, low-magnetic field regions.

The time that a particle can travel before either the turbulent
component of the magnetic field changes or the particle gets absorbed
is
\begin{equation}\label{tdist}
t_{\rm step}=-\ln\left(\frac{\chi}{c (k_{\rm sc}+k_{\rm abs})}\right)\enspace,
\end{equation}
where $\chi$ is a standard deviate sampled between 0 and 1. This comes
from the exponential nature of both the absorption and magnetic field
coherence length with distance travelled by the particle. 

When a particle's trajectory would take it out of its current cell and
into the next cell, the code instead scales back the particle's
position so as to just barely enter the next cell. In doing this,
there is an implicit assumption on the particle steps being drawn from
the same distribution, even though in practice some of those steps are
shorter than probability would suggest. However, this should not
strongly affect the results of the simulation, especially in the limit
that the particle takes many steps before exiting each cell.

Within this region of coherent magnetic field, the particle propagates
in a way which is dramatically different from diffusion. The particle
is strongly biased to move in a single direction, along the coherent
magnetic field, and travels in that direction quickly, with a
mean-free-path which is proportional to the time-of-flight, as opposed
to the square root of the time-of-flight which diffusion does.
\subsection{Discrete Diffusion Monte Carlo}
In regions where there is no coherent magnetic field, the chaotic
magnetic fields change direction on a scale much smaller than the grid
spacing of the code. The particles require many small, random-walk
steps to reach the edge of a cell-wall. In these regions, a diffusion
approximation is valid. To calculate the behavior of a cosmic ray in
the diffusive regime, we use discrete diffusion Monte Carlo (DDMC)
methods. In DDMC, many of the smaller random-walk steps which would be
taken by the particle are combined together into larger groups of
steps. These steps are drawn at random from a distribution of particle
locations after $N$ steps. Specifically, we choose $N$ such that it
will take several of these DDMC steps to leave a given cell. However,
this speeds up the code over direct Monte Carlo calculation by many
orders of magnitude. Effectively, the code recalculates the diffusion
coefficient in each cell, according to the equations derived below. This allows the geometry to, in principle,
have a different turbulent field amplitude in each cell, though in
this study each cell outside of the coherent field region the same
turbulent field amplitude. Inside the coherent field region, for this
study, we assume the turbulent field amplitude is zero.

The particle diffusion is done according to a standard
three-dimensional random walk according to the diffusion coefficient
within each cell. To calculate the diffusion coefficient in
each cell, we ran our particle transport Monte Carlo for many
particles, on the assumption that even turbulent magnetic fields are
coherent on distances much shorter than their coherence length. We
then averaged over those particles' motion according to the equations
described below, in order to calculate the appropriate diffusion
behavior in each cell according to the strength of its turbulent
field.

While the properties of coherent magnetic field $\vec{B}_{\rm coh}$
have their amplitude and direction fixed by the input geometry, the
turbulent magnetic field $\vec{B}_{\rm turb}$ varies step-by-step
throughout the propagation. For the turbulent magnetic field, we
assume that each $\vec{B}_{\rm turb}$ value is sampled at
random. (This is the real-space analog of a Kolmogorov spectrum). Note
that this is different from sampling a random {\it amplitude} of the
turbulent magnetic field from a flat distribution. For a random vector
in three dimensions $(B_{\rm turb},\theta,\phi)$ the amplitude $B_{\rm
  turb}$, polar angle $\theta$, and azimuthal angle $\phi$ are sampled
as
\begin{eqnarray}
B_{\rm turb}&=&b_{\rm t}\chi_1^{1/3}\label{bamp}\\
\cos(\theta)&=&2\chi_2-1\label{costheta}\\
\phi&=&2\pi\chi_3\label{phi}\enspace,
\end{eqnarray}
with $\chi_i$ standard deviates between 0 and 1. The maximum allowed
value of the turbulent field amplitude $b_{\rm t}$ sets the scale for
allowed values of the magnetic field. This gives a mean amplitude of
the turbulent magnetic field of $\langle B_{\rm turb}\rangle=0.75b_{\rm t}$.

For a three-dimensional random walk with typical distance per step
$a$, the probability of moving distance $r$ after $N$ steps is the
three-dimensional Gaussian
\begin{equation}\label{probr}
P(r;N)=4\pi r^2\left(\frac{3}{2\pi N a^2}\right)^{3/2}\exp\left(-\frac{3r^2}{2N a^2}\right)\enspace.
\end{equation}
This can be most easily calculated by sampling three one-dimensional
Gaussian random variables $g_1, g_2, g_3$ centered at 0 with widths
$\sqrt{N a^2/3}$ and adding them in quadrature:
\begin{equation}
g_{3D}=\sqrt{g_1^2+g_2^2+g_3^2}\enspace,
\end{equation}
where Gaussian random variables are calculated using, e.g., a
Box-Muller transform. The direction is then given using a polar and
azimuthal angle sampled as in equations~\ref{costheta} and~\ref{phi}.

The typical distance for each step follows a similar equation for each
step of the transport routine (equation~\ref{tdist}), but averaged
over many steps. Because the magnetic field changes in each step,
thereby changing the value of $K_{\rm sc}$, this must be taken into
account as well. The correct value of $a$ is the expectation value for
displacement in a single step, given by
\begin{equation}\label{rms}
a=\sqrt{\langle\vec{r}^2\rangle-\langle\vec{r}\rangle^2}=\sqrt{\langle\vec{r}^2\rangle}\enspace.
\end{equation}
For simplicity, we assume that the energy change of the particle over
a step is negligible. For convenience, we also define a dimensionless
parameter
\begin{equation}
\xi\equiv \frac{k_{\rm sc}(B_{\rm t}=b_{\rm t})}{k_{\rm abs}}\enspace,
\end{equation}
where $k_{\rm sc}$ (equation~\ref{ksc}) is evaluated at the maximum
turbulent magnetic field $b_{\rm t}$ and $k_{\rm abs}$ is calculated
as in equation~\ref{kabs}.  For the magnetic field of
equation~\ref{bamp}, the expectation value of the step size is
\begin{align}\label{rmsd}
a=\frac{\sqrt{2}}{k_{\rm abs}}&\left[\frac{3}{2}\xi^{-2}-4\xi^{-3}+9\xi^{-4}-24\xi^{-5}+6\frac{\xi^{-6}}{1+\xi}\right.\nonumber\\
&\left. +30\xi^{-6}\ln(1+\xi)-6\xi^{-6}\right]^{1/2}\enspace.
\end{align}
In the limiting cases of large $\xi$ ($k_{\rm sc}\gg k_{\rm abs}$) and
small $\xi$ ($k_{\rm abs}\gg k_{\rm sc}$), this simplifies to
\begin{eqnarray}
a\approx\frac{\sqrt{3}}{k_{\rm sc}(B_{\rm t}=b_{\rm t})}, && \xi\rightarrow\infty\\
a\approx\frac{\sqrt{2}}{k_{\rm abs}}, && \xi\rightarrow0\enspace.
\end{eqnarray}
The value of $a$, in
equation~\ref{probr}, determines the typical displacement of the
particle after $N$ steps.

To determine whether the particle was absorbed into the ISM after
these $N$ steps, we must determine the amount of time that the
propagation took, or equivalently, the path-length of the particle's
motion. Rather than the values $\langle\vec{r}^2\rangle$ and
$\langle\vec{r}\rangle$ which went into equation~\ref{rms}, the path
length is evaluated as $\langle\lvert\vec{r}\rvert\rangle$. Following
a similar calculation to equation~\ref{rmsd}, we find that the typical
path length for each step is
\begin{align}
\langle\lvert\vec{r}\rvert\rangle=\frac{1}{k_{\rm abs}}&\left[\frac{6}{5}\xi^{-1}-\frac{3}{2}\xi^{-2}+2\xi^{-3}-3\xi^{-4}+6\xi^{-5}\right.\nonumber\\
&\left.-6\xi^{-6}\ln(1+\xi)\right]\enspace.
\end{align}
In the limiting cases of large $\xi$ ($k_{\rm sc}\gg k_{\rm abs}$) and
small $\xi$ ($k_{\rm abs}\gg k_{\rm sc}$), this simplifies to
\begin{eqnarray}
\langle\lvert\vec{r}\rvert\rangle\approx\frac{6}{5k_{\rm sc}(B_{\rm t}=b_{\rm t})}, && \xi\rightarrow\infty\\
\langle\lvert\vec{r}\rvert\rangle\approx\frac{1}{k_{\rm abs}}, && \xi\rightarrow0\enspace.
\end{eqnarray}
The probability that the particle was absorbed by the ISM is
\begin{equation}
P_{\rm abs}(N)=\exp(- k_{\rm abs} N \langle\lvert\vec{r}\rvert\rangle)
\end{equation}
and the time taken for the N steps is 
\begin{equation}
\Delta t_N=N\langle\lvert\vec{r}\rvert\rangle/c\enspace.
\end{equation}

When a particle's trajectory would take it out of its cell and into
another, the trajectory is scaled back, similar to what is done during
the particle transport routine discussed in
section~\ref{PTMC}. However, because the displacement is proportional
to the square root of the number of steps, the number of steps is
scaled back quadratically to get the propagation distance and time. To
propagate over the distance from the source to the boundary wall
($75-130{\rm\,pc}$) takes a typical particle 4-20 million years.

\subsection{Ion Energy Losses}
After each time step in the transport and diffusion processes, we
calculate the energy losses for the particle. For relativistic
nucleons in a charged plasma, such as the ISM, the energy losses are
dominated by Coulomb scattering and ionization of the medium. We
calculate the energy losses for ions in our simulation as a continuous
process, following~\cite{Strong:1998pw}. Energy losses from
Coulomb collisions are given by~(\cite{Mannheim:1994sv})
\begin{equation}
\left(\frac{\mathrm{d}E}{\mathrm{d}t}\right)_{\rm Coulomb}\approx -4 \pi r_e^2 c m_e c^2 Z^2 n_e \ln\Lambda \frac{\beta^2}{x_m^3+\beta^3}\enspace,
\end{equation}
where $r_e=2.8\times10^{-13}{\rm\,cm}$ is the classical electron
radius, $m_e c^2=0.511{\rm\,MeV}$ is the electron mass, $Z e$ is
the charge of the incoming ion, $\beta=v/c$ is the speed of the
incoming ion, and $n_e$ is the number density of electrons in the
plasma. The temperature of the electrons in the plasma, $T_e$, comes
in as
\begin{equation}
x_m\equiv\left(\frac{3\sqrt{\pi}}{4}\right)^{1/3}\sqrt{\frac{2 k_B T_e}{m_e c^2}}\enspace,
\end{equation}
with $k_B$ the Boltzmann constant. The Coulomb logarithm $\ln\Lambda$
accounts for all possible scattering angles of the incident ion with
the electron~(\cite{1985ApJ...295...28D}). Because the incoming ion and
the plasma electron are non-identical particles, the maximum
scattering angle is $\pi$ (for identical particles, it would be
$\pi/2$). The minimum scattering angle is given by the excitation of a
plasmon with the plasma frequency. For a cold plasma in the Born
regime, which is appropriate for the ISM, this is given in equation B7
of~\cite{1985ApJ...295...28D}. The corresponding Coulomb logarithm is
\begin{equation}
\ln\Lambda\approx\frac{1}{2}\ln\left(\frac{m_e^2 c^4}{\pi r_e \hbar^2 c^2 n_e}\frac{M \gamma^2 \beta^4}{M+2 \gamma m_e}\right)\enspace,
\end{equation}
where $\hbar$ is the reduced Planck constant, $M$ is the incoming ion
mass, and $\gamma$ is the incoming ion Lorentz factor. Note that in
calculating the energy losses, we account for the true speed of the
incoming ion, using
\begin{eqnarray}
\gamma&=&\frac{E}{M c^2}\\
\beta&=&\sqrt{1-\frac{1}{\gamma^2}}\enspace.
\end{eqnarray}

The other primary source of energy loss for relativistic ions is
through ionization of the neutral plasma. This follows the Bethe-Bloch
formula~(\cite{Strong:1998pw,Mannheim:1994sv}):
\begin{eqnarray}
Q_{\rm max}&=&\frac{2m_ec^2\beta^2\gamma^2}{1+2\gamma\frac{m_e}{M}+\frac{m_e^2}{M^2}}\\
B_{\rm H}&=&\ln\left[\frac{2m_e c^2\beta^2\gamma^2 Q_{\rm max}}{(19{\rm\,eV})^2}\right]-2\beta^2\\
B_{\rm He}&=&\ln\left[\frac{2m_ec^2\beta^2\gamma^2Q_{\rm max}}{(44{\rm\,eV})^2}\right]-2\beta^2\\
\left(\frac{\mathrm{d}E}{\mathrm{d}t}\right)_{\rm Ionization}&=&-2 \pi r_e^2 c m_e c^2 Z^2 \frac{1}{\beta} n \nonumber\\
&\times&(x_{\rm H}B_{\rm H}+2x_{\rm He}B_{\rm He})\enspace,
\end{eqnarray}
where $Q_{\rm max}$ is the kinematic maximum energy transfer to the
atomic electron, $19{\rm\,eV}$ is the ionization energy of neutral
hydrogen, $44{\rm\,eV}$ is the ionization energy of neutral helium,
$n$ is the combined number density of neutral hydrogen and helium, $x_{\rm H}$ and $x_{\rm He}$ are the fraction of the neutral gas made up of hydrogen and helium, and
we have accounted for the two ionizable electrons on the helium
atoms. Because $M>>m_e$, we neglect the quadratic term in $Q_{\rm
  max}$ in our calculations. Other energy losses are negligible for
highly relativistic ions. For proton ions travelling through the ISM,
ion fragmentation and decay are not considered.
\section{Results}\label{results}
\begin{table}
\begin{center}
\begin{tabular}[t]{|c|c|c|c|c|}
  \hline
  model designation & A & B & C & D \\
  \hline
  $a_{\rm min}$ (pc) & 111 & 111 & 81 & 138 \\
  \hline
  $a_{\rm max}$ (pc) & 120 & 114 & 84 & 141 \\
  \hline
  distance from source (pc) & 57.3 & 53.0 & 10.6 & 91.2 \\
  \hline
\end{tabular}
\caption{ Description of the models of coherent magnetic field. The
  coherent field is a region which is $a_{\rm min}<y,z<a_{\rm max}$
  and points in the x-direction (toward Earth). The distance from the
  source to center of each coherent magnetic field is shown as well.
\label{BfieldModels}}
\end{center}
\end{table}
\begin{table}
\begin{center}
\begin{tabular}[t]{|c|c|c|c|c|}
  \hline
  model designation & A & B & C & D \\
  \hline
  particle density & $1.6$ & $1.0$ & $2.8$ & $2.9$ \\
  at Earth (prot/sr)&$\times10^7$&$\times10^8$&$\times10^8$&$\times10^7$\\
  \hline
  particle density, & $1.5$ & $1.8$ & $2.6$ & $1.9$ \\
  no coherent field (prot/sr)&$\times10^5$&$\times10^5$&$\times10^5$&$\times10^4$\\
  \hline
  increased brightness over& $110\times$ & $560\times$ & $1080\times$ & $1500\times$ \\
  no coherent field &&&&\\
  \hline
\end{tabular}
\caption{ The details of the particle density observed in the sky
  toward the magnetic field region for each considered geometry. The
  simulation results were calculated from $2\times10^6$ sampled 10\,TeV
  protons. The models of coherent magnetic field in the first row are
  given in table~\ref{BfieldModels}. The second row gives the particle
  density (particles/sr) which arrives at the $x=150{\rm\,pc}$ wall,
  which is our proxy for the Earth sky. The third row gives the
  particle density at that same location in a simulation with no
  coherent magnetic field, for comparison. The ratio of the particle
  density with and without the coherent field is given in row
  four. (See section~\ref{Pdens} for more details.)
\label{PdensTable}}
\end{center}
\end{table}
\begin{table}
\begin{center}
\begin{tabular}[t]{|c|c|c|c|c|}
  \hline
  model designation & A & B & C & D \\
  \hline
  fraction particles& 12\% & 9.0\% & 46\% & 1.2\% \\
  within coherent field &&&&\\
  \hline
  geometric $\Omega/4\pi$ & 2.8\% & 0.93\% & 5.7\% & 0.41\% \\
  of coherent field&&&&\\
  \hline
  increased particles& $4.3\times$ & $9.7\times$ & $8.1\times$ & $2.9\times$ \\
  trapped from scattering &&&&\\
  \hline
\end{tabular}
\caption{ The relative importance of particles which scatter into the
  magnetic field compared to just the geometric solid angle it
  subtends from the source. The simulation results were calculated
  from $2\times10^6$ sampled 10\,TeV protons. The models of coherent
  magnetic field in the first row are given in
  table~\ref{BfieldModels}. Row two gives the fraction of the
  isotropic flux which is absorbed into the coherent magnetic
  field. Row three gives the fraction of the total $4\pi$ solid angle
  which is subtended by the coherent magnetic field. Row four gives
  the ratio of the total acceptance of particles into the coherent
  region to the geometric solid angle. This in indicative of the
  importance of particles which scatter into the coherent field.  (See
  section~\ref{acceptance} for more details.)
\label{acceptanceTable}}
\end{center}
\end{table}
\begin{figure}[t]
\begin{center}$
\begin{array}{cc}
\includegraphics[width=0.5\columnwidth]{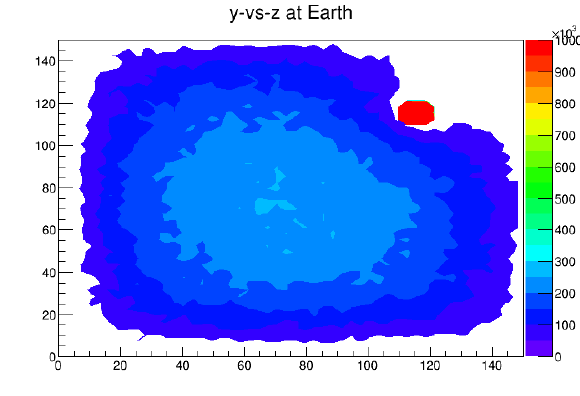} & \includegraphics[width=0.5\columnwidth]{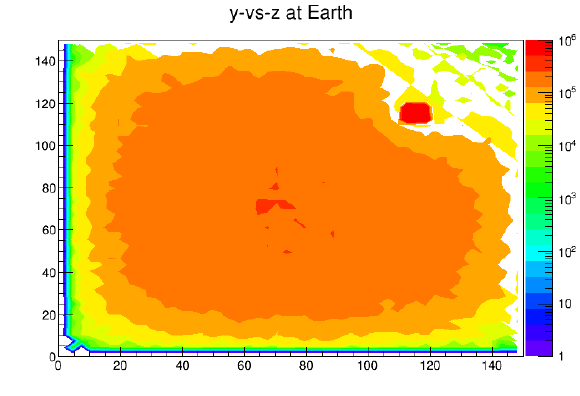}\\
\includegraphics[width=0.5\columnwidth]{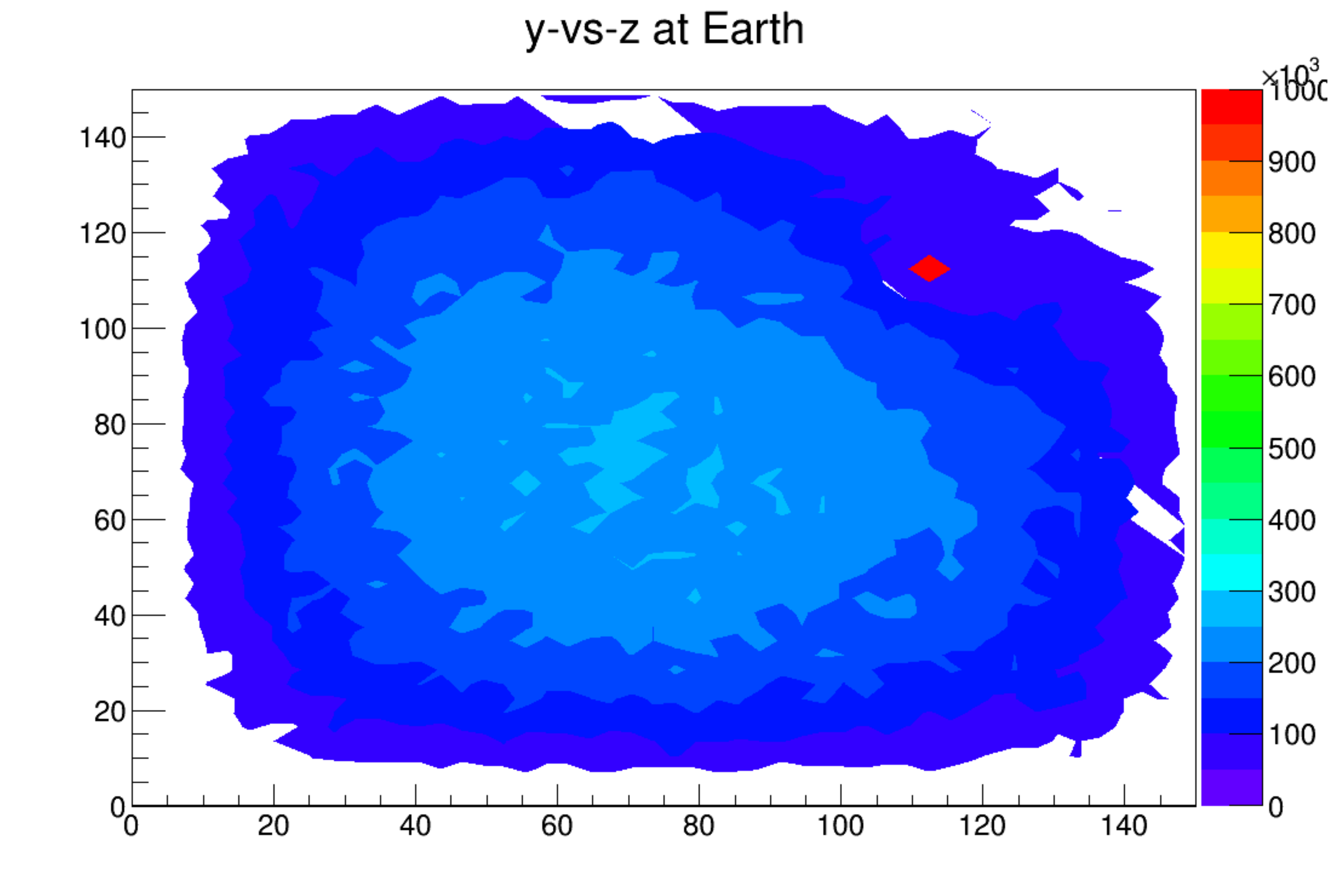} & \includegraphics[width=0.5\columnwidth]{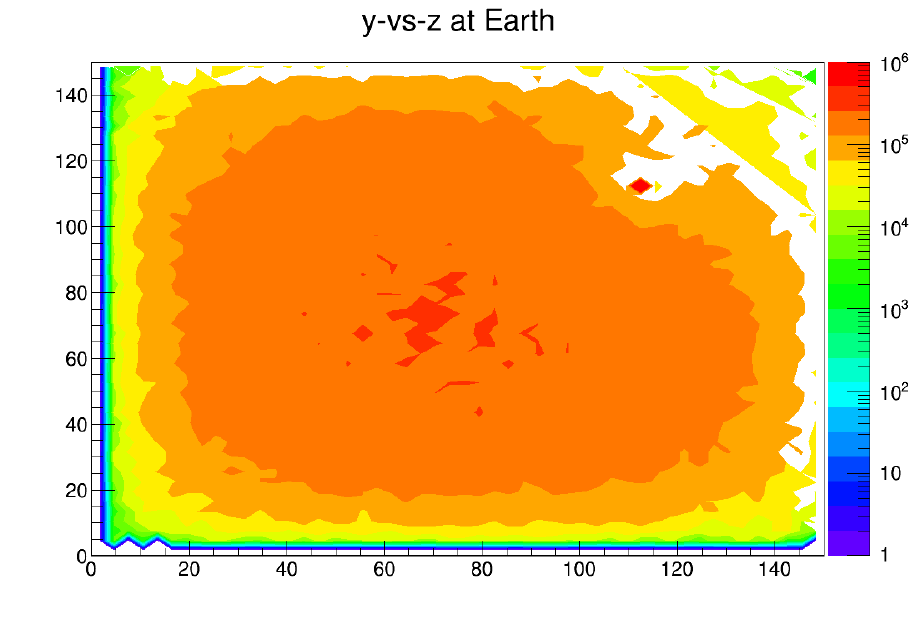}
\end{array}$
\end{center}
\caption{\small A comparison of models A (top) and B (bottom) of
  table~\ref{BfieldModels}. This compares the effect of physical
  extent of the region of coherent magnetic field on particle
  propagation. The axes are y and z at the $x=150{\rm\,pc}$ wall,
  which is our proxy for the Earth sky, and the color scale is in
  particles/sr, shown both linearly (left) and logarithmically
  (right). The plots were made with $2\times10^6$ sampled 10\,TeV
  protons. The bright red spot in the upper right quadrant of the
  figures is the location of the coherent magnetic field. Of
  particular note is the region with a deficit of particles
  surrounding the coherent magnetic field and to its upper right. This
  is caused by the high probability of particles which would enter
  these regions first scattering into the coherent magnetic field. For
  more on this, see section~\ref{SizeB}.
\label{bigvssmall}}
\end{figure}
\begin{figure}[ht]
\begin{center}$
\begin{array}{cc}
\includegraphics[width=0.5\columnwidth]{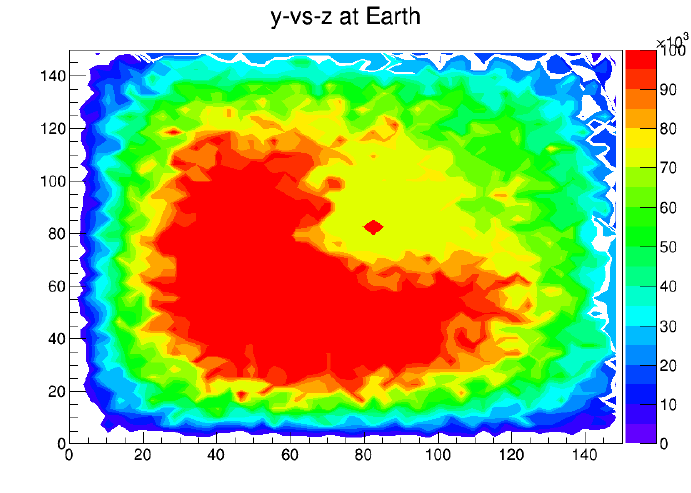} & \includegraphics[width=0.5\columnwidth]{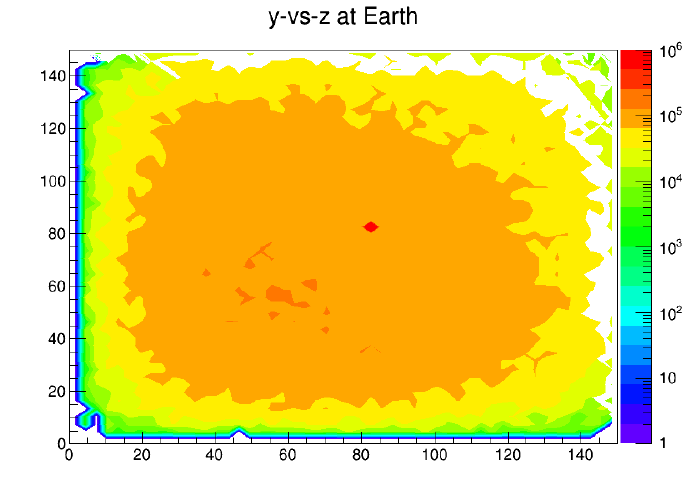}\\
\includegraphics[width=0.5\columnwidth]{2Mparticles_111-114Bfield_linscale.pdf} & \includegraphics[width=0.5\columnwidth]{2Mparticles_111-114Bfield_logscale.pdf}\\
\includegraphics[width=0.5\columnwidth]{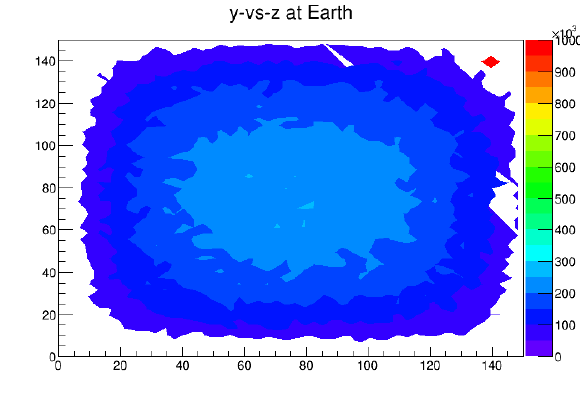} & \includegraphics[width=0.5\columnwidth]{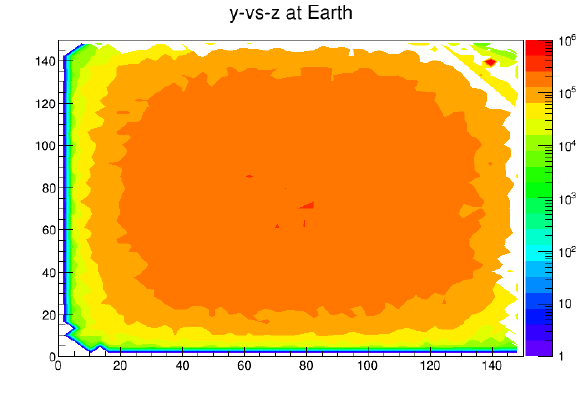}
\end{array}$
\end{center}
\caption{\small A comparison of models C (top), B (middle), and D
  (bottom) of table~\ref{BfieldModels}. This compares the effect of
  distance from the source to the region of coherent magnetic field on
  particle propagation. The axes are y and z at the $x=150{\rm\,pc}$
  wall, which is our proxy for the Earth sky, and the color scale is
  in particles/sr, shown both linearly (left) and logarithmically
  (right). Note that the linear-scale pot for model C only goes up to
  $10^5$, unlike the rest of the figure which have color scales up to
  $10^6$. (This was done to highlight the deficit region in this
  figure. On a scale up to $10^6$, this figure has a bright red spot
  on a monochrome blue background.) The plots were made with
  $2\times10^6$ sampled 10\,TeV protons. The bright red spot in the
  upper right quadrant of the figures is the location of the coherent
  magnetic field. Similar to figure~\ref{bigvssmall}, there is a
  region with a deficit of particles surrounding the coherent magnetic
  field and to its upper right. This effect becomes much stronger as
  the coherent magnetic field is moved closer to the source. For more
  on this, see section~\ref{DistB}.
\label{changingdistance}}
\end{figure}
To test how a region of coherent magnetic field would affect
propagating cosmic rays, we have run several simulations with varying
locations and spatial sizes of coherent magnetic fields. All the tests
were run within a $150{\rm\,pc}\times150{\rm\,pc}\times150{\rm\,pc}$ box, with a grid
spacing of $3{\rm\,pc}$. For all of the tests, we assume a source located
at the center of the box (at (75\,pc,75\,pc,75\,pc)) emitting
monochromatic particles with an energy of 10\,TeV. The simulated
cosmic rays are 100\% protons. All physical parameters of the ISM are
given in table~\ref{paramstab}. We assume a region of coherent
magnetic field which is uniform in the x-direction and contained to a
finite region $a_{\rm min}<y,z<a_{\rm max}$ in the y- and
z-directions. The specific magnetic field regions considered are
listed in table~\ref{BfieldModels}. For the parameters modeled, the
energy losses of the protons were small ($\sim1-10{\rm\,MeV}$) and did not appreciably affect
the particles' energies.

Our proxy for the anisotropies seen in the sky at Earth is the
$x=150{\rm\,pc}$ wall of our simulation. This models the case where a
region of coherent magnetic field stretches from the general vicinity
of a cosmic-ray source to dump its particles near to the Earth. We
consider the relative particle density per solid angle, which
corresponds to the signals observed over given solid angle in the sky
from Earth.

This study is designed to show the
plausibility of such coherent magnetic fields to reproduce anisotropic
cosmic-ray features, rather than a full study of how the magnetic
field properties affect such features. Therefore, we do not include
magnetic field feedback from the cosmic rays or turbulent magnetic
fields within the coherent magnetic field region. These effects will
be included in a follow-up publication~(\cite{HardingFryer}).

\subsection{Increased Particle Density within the Coherent Field}\label{Pdens}
While the particles' diffusion causes a smooth, dipole-like
distribution on the sky, the particles that get trapped within the
region of coherent magnetic field give a much sharper feature on the
sky. In this region, the transport of the cosmic rays follows
$\vec{v}\times\vec{B}$ and the particles move along the magnetic field
quickly, moving linearly with time. Because of this, the particles
which enter the region of coherent field tend to not leave the region
during their propagation, until they reach the boundary of the
simulation. We expect that this is still true if there are turbulent
fields within this region, since turbulent diffusion is a slower
process, proportional to the square root of propagation time. 

With the greater number of particles being deposited within a small
solid angle, the flux from such a region is much larger than from a
purely diffusive simulation. Compared to the flux expected from pure
diffusion, the flux from the coherent magnetic field is increased by
2-3 orders of magnitude. Specific numbers for our four magnetic field
models can be found in table~\ref{PdensTable}.

\subsection{Generality of the Coherent Magnetic Field Location}
One notable feature of the coherent magnetic fields is that the
magnetic field does not have to be directly lined up with a source to
still give a large flux on the sky. Previously, the idea of coherent
magnetic fields causing anisotropies in the sky were
considered~(\cite{Drury:2008ns}), but it was assumed that these magnetic
fields needed to act as a bridge to channel particles between a source
and the Earth. However, it seems that these coherent fields act more
as a superhighway, moving the particles from one place to another but
not necessarily directly aligned with the source itself. The effect is
to move the cosmic rays to a location much closer to the Earth,
creating in effect a new, closer cosmic-ray source which may have
large angular extent and is not necessarily coincident with photon or
neutrino signatures of a true astrophysical source. These behaviors
can be seen in figures~\ref{bigvssmall} and~\ref{changingdistance}.

\subsection{Scattering Increases the Importance of the Coherent Field}\label{acceptance}
Another feature of the coherent magnetic fields is that they actually
contain more cosmic rays than one might expect. Because the coherent
magnetic field covers a region of the sky through which the source's
particles must pass, the expected number of particles within the
coherent field could be expected to be proportional to the solid angle
from the source subtended by the coherent field. However, when we
simulate the cosmic-ray propagation, we find that far more particles
are contained within the coherent field than can be explained with
geometric solid angle alone. These numbers for our considered
geometries are given in table~\ref{acceptanceTable}.

In addition to the particles which would be caught in the coherent
field while traveling radially outward, particles scattering in the
turbulent magnetic field will also often scatter into the coherent
magnetic field region during their random walk. Once inside the
coherent field, these particles are transported quickly along the
field and cease their random walk. Therefore, these scattered
particles increase the number trapped by the coherent field over just
a geometric value. Depending on the model, this increase can be 3-10
times the geometric area, and is typically a more important effect for
smaller coherent field regions. In figures~\ref{bigvssmall}
and~\ref{changingdistance}, the lack of particles which scatter past
the coherent magnetic field without passing through it can be seen as
a deficit in the distribution in the sky to the upper right of the
coherent magnetic field and in its general vicinity.

And intriguing feature of this effect is that as the turbulent fields
become stronger, the importance of coherent fields becomes stronger as
well, even if the magnetic field strength of the coherent field does
not change. So in regions where coherent fields are a {\it lesser}
part of the energy budget, the coherent fields are {\it more}
important for particle propagation behavior.

\subsection{The Spatial Size of the Field}\label{SizeB}
Figure~\ref{bigvssmall} shows the sky from two of our coherent
magnetic field models - model A and B of
table~\ref{BfieldModels}. These two models are at similar distances
from the source, with the primary difference being their width - model
A is $9\times9$\,pc while model B is $3\times3$\,pc. As is shown in
table~\ref{acceptanceTable}, the coherent magnetic field region of
model A contains more particles than the small region of model
B. However, model B only contains 25\% fewer particles than model B,
despite model B having a cross-sectional area which is nine times
larger. Therefore, the density of particles per solid angle on the sky
actually {\it increases} by nearly an order of magnitude for {\it
  smaller} regions of coherent magnetic field.

\subsection{The Distance from the Source to the Field}\label{DistB}
The final effect we studied was how the distance from the source to
the coherent magnetic field affected the number of particles trapped
within the coherent field. These results, for models B, C, and D, are
shown in figure~\ref{changingdistance}. As expected, the number of
particles trapped by the coherent magnetic field is largest for models
closer to the source, with model C at a distance of 10\,pc containing
nearly 50\% of the total particles output by the source. However, the
expected particle density from models with no coherent magnetic field
also decreases slightly as you look at sky positions far from the
source, and the solid angle for locations near the edge of the wall is
slightly smaller than that near the center of the wall. Therefore,
although the number of particles within the coherent field decreases
by nearly forty times between model C and D
(table~\ref{acceptanceTable}), the particle density in the sky only
changes by an order of magnitude. Also, the relative improvement from
the coherent magnetic field over no coherent field actually rises
(table~\ref{PdensTable}) when the coherent field is very far from the
source. For a coherent magnetic field this far from the source to have
such a large improvement in particle flux over expected, it is
possible that such features are important for diffuse, isotropic
cosmic rays as well.
\begin{figure*}[t]
\begin{center}$
\begin{array}{ccc}
\includegraphics[width=0.5\columnwidth]{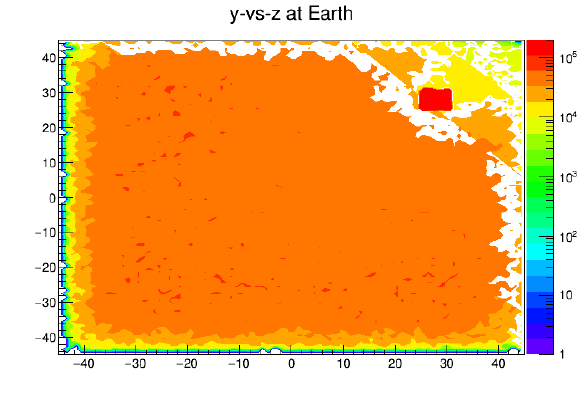} & 
\includegraphics[width=0.5\columnwidth]{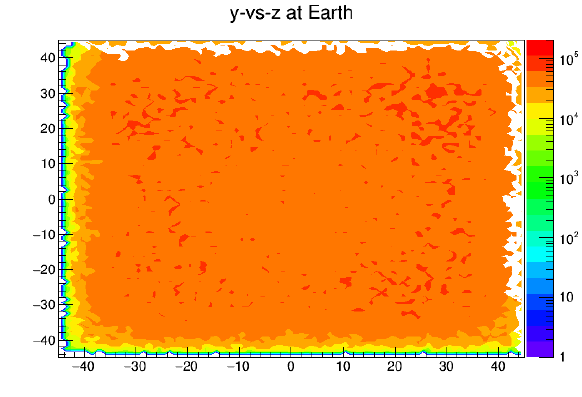} &
\includegraphics[width=0.5\columnwidth]{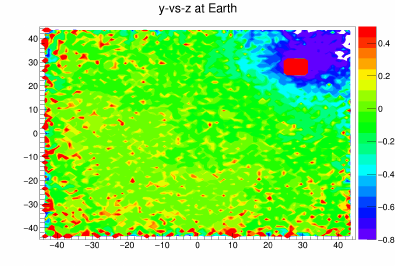}
\end{array}$
\end{center}
\caption{\small Skymaps, in terms of angle from the CR source. On the
  left is a skymap for model A of table~\ref{BfieldModels}. The plots
  were made with $2\times10^6$ sampled 10\,TeV protons. The angular
  scale for such a large magnetic field region is roughly
  $5\degree\times5\degree$ in extent. In the center is a skymap which
  was run with only diffusive propagation and no coherent
  regions. Fluctuations in this map are dominated by the finite
  statistics of the two million particles run in the simulation and
  would decrease as the number of particles is increased. On the right
  is a relative intensity map between the two other figures, which
  shows that the coherent field is easily distinguished above
  background fluctuations (the region actually has a relative
  intensity of roughly 25). 
\label{relint}}
\end{figure*}

\section{Outlook}

These studies are meant as a first look at how coherent magnetic
fields in the nearby ISM can cause anisotropic signals of cosmic
rays. In particular, we have shown that such signals exist even with
general coherent magnetic field configurations - they do not have to
contain a cosmic-ray source. Coherent magnetic fields can increase the
flux of particles in their direction in the sky by greater than three
orders of magnitude, and even coherent fields far from a cosmic-ray
source can contain several percent of the source's emitted
particles. These effects get stronger for smaller regions of coherent
field and for regions with more turbulent diffusion. The less of the
energy budget that is in coherent fields, the more important they are
to propagation.

Other studies of the cosmic-ray anisotropies have seen similar results
to those in section~\ref{results}. In both~\cite{Ahlers:2013ima}
and~\cite{Ahlers:2015dwa}, it was shown that turbulent structures in
the local magnetic field can produce small-scale anisotropies through
anisotropic diffusion. \cite{Giacinti:2011mz} also showed that these
local magnetic field structures could create such
anisotropies. Similar work to our code was calculated in
\cite{Lopez-Barquero:2015qpa}. However, that study primarily looked at
particles with PeV-scale energies, where the Larmor radius and
mean-free-path are on parsec scales, similar in size to the coherence
length of the magnetic turbulence. An integral part of our code
presented in this work was to consider lower-energy TeV-scale
particles which have Larmor radii and mean-free-paths which are much
smaller than coherence length of the structured magnetic fields. Full
MC integration of these TeV particles is intractably slow, so this was
why a hybrid diffusion+MC method was needed. Additionally,
\cite{Lopez-Barquero:2015qpa} assumed that the source population was
diffuse, which allowed them to time-reverse incoming particles to get
maps of the anisotropic sky. However, our code allows for general
source distributions in which particular CR sources are dominating the
anisotropy.

Due to the inclusion of anisotropic cosmic rays sources, our code
cannot conveniently be time-reversed to create artificial
"back-tracked" skymaps as was done in Lopez-Barquero et al
2015. Instead, we used the method suggested in that paper of
increasing the size of the target to record those particles that pass
"nearby". Using this method, we can convert from our earth-proxy maps
(in pc) to angular maps of the Earth sky. However, because of low
statistics, we can only get a reasonable sample out to within 45 degrees
of the source location.

We have made such maps, both in the case of our model A and in the
case of no coherent magnetic fields. The relative intensity of the
magnetic field geometry over a geometry without coherent fields at
each angle is shown in figure~\ref{relint}. As can be seen in that
figure, the strength of the anisotropy in our geometry A is large,
much larger than is currently observed at Earth. However, it is
expected that this intensity can be affected by the details of the
magnetic field strength and geometry and the source spectra and
geometry. However, this early result indicates that similar relative
intensity features to the observed anisotropy should be possible.

Further studies of the cosmic-ray anisotropies with anisotropic
transport through coherent magnetic fields may yield more information
about how these anisotropies behave. In addition to isolated regions
of cosmic-ray excess, some excess regions are also seen as long, thin
regions extending in declination. Also, the location of the cosmic-ray
anisotropy seems to vary with cosmic-ray energy
~(\cite{Aartsen:2012ma}).  A more detailed study of these features may
suggest details about the structure of the magnetic fields which are
causing them.  It is also possible that the interplay between coherent
magnetic fields and turbulent magnetic fields may be related to these
additional features of the anisotropy~(\cite{HardingFryer}).

The interaction between coherent and turbulent magnetic fields could
also be important due to cross-fieldline transport for structures with
narrow geometrical structures. In this case, the ``beaming'' effects
we observed here may be somewhat smeared out and less significant. In
future work with this code, we plan on addressing such issues,
including having both coherent and turbulent fields in the same
spatial regions. With the inclusion of such features, the code will be
more broadly applicable and the effects of coherent field structures
on the anisotropy will be better understood.

It is possible that some of the features of the observed anisotropy
are coincident with the heliotail~(\cite{Drury:2008ns,Lazarian:2010sq}),
which is a known region of nearby high magnetic field. Also, IBEX has
recently begun surveying the magnetic field structure of the nearby
galaxy. These structures, along with the observed cosmic-ray
anisotropies, may be able to explain the local cosmic-ray flux and
begin to study the sources which create these observed cosmic rays. We
will consider this is a forthcoming publication~(\cite{HardingFryer}).
Depending on the details of the local ISM, non-diffusive propagation
can explain the observed TeV cosmic-ray anisotropy.

\begin{acknowledgments}
We thank Brenda Dingus and Fan Guo for useful discussions. Work at
LANL was done under the auspices of the National Nuclear Security
Administration of the U.S. Department of Energy at Los Alamos National
Laboratory under Contract No. DE-AC52-06NA25396 through an IGPPS
grant.
\end{acknowledgments}

\bibliography{bibliography}
\end{document}